\documentclass{article}
\usepackage{ijcai25}

\usepackage{times}
\usepackage{soul}
\usepackage{url}
\usepackage[hidelinks]{hyperref}
\usepackage[utf8]{inputenc}
\usepackage[small]{caption}
\usepackage{graphicx}
\usepackage{amsmath}
\usepackage{amsthm}
\usepackage{booktabs}
\usepackage{algorithm}
\usepackage{algorithmic}
\usepackage[switch]{lineno}
\usepackage{makecell}
\usepackage{multirow}
\usepackage{amssymb}

\AtBeginDocument{%
  }

\title{Hand by Hand: LLM Driving EMS Assistant for Operational Skill Learning}

\author{
Wei Xiang
\and
Ziyue Lei
\and
Haoyuan Che
\and
Fangyuan Ye
\and
Xueting Wu
\And
Lingyun Sun\\
\affiliations
Zhejiang University, Hangzhou, China\\
\emails
\{wxiang, zy\_lei, chaoyuan0116\}@zju.edu.cn,
yefangyuanzju@gmail.com,
3235577560@qq.com,
sunly@zju.edu.cn
}
\begin{document}

\maketitle
\begin{abstract}
    Operational skill learning, inherently physical and reliant on hands-on practice and kinesthetic feedback, has yet to be effectively replicated in large language model (LLM)-supported training. Current LLM training assistants primarily generate customized textual feedback, neglecting the crucial kinesthetic modality. This gap derives from the textual and uncertain nature of LLMs, compounded by concerns on user acceptance of LLM driven body control. To bridge this gap and realize the potential of collaborative human-LLM action, this work explores human experience of LLM driven kinesthetic assistance. Specifically, we introduced an ``Align-Analyze-Adjust'' strategy and developed FlightAxis, a tool that integrates LLM with Electrical Muscle Stimulation (EMS) for flight skill acquisition, a representative operational skill domain. FlightAxis learns flight skills from manuals and guides forearm movements during simulated flight tasks. Our results demonstrate high user acceptance of LLM-mediated body control and significantly reduced task completion times. Crucially, trainees reported that this kinesthetic assistance enhanced their awareness of operation flaws and fostered increased engagement in the training process, rather than relieving perceived load. This work demonstrated the potential of kinesthetic LLM training in operational skill acquisition.
\end{abstract}

\section{Introduction}
%需要表现出技能训练的重要性
Operational skills such as driving, riding, and operating surgeries occupied diverse aspects of life and work. Mastering operational skills requires not only theoretical knowledge but also procedure precision and operation stability, ensuring critical metrics within acceptable ranges \cite{Feedback}. While traditional training methods, such as self-directed reflection and expert instruction, remain valuable, Large Language Model (LLM)-supported assistance has recently emerged as a promising and rapidly expanding area of research. LLM acquires expertise and experience from extensive literature, and possesses the ability to analyze multi-modal training data, positioning them to provide tailored feedback aligned with trainees' progress \cite{ijcai2024p1045}. Furthermore, LLM-supported systems offer personalized guidance and professional responses, supporting trainees to practice at an individualized pace \cite{ijcai2024p728}. 
\begin{figure}
    \centering
    \includegraphics[width=1.0\linewidth]{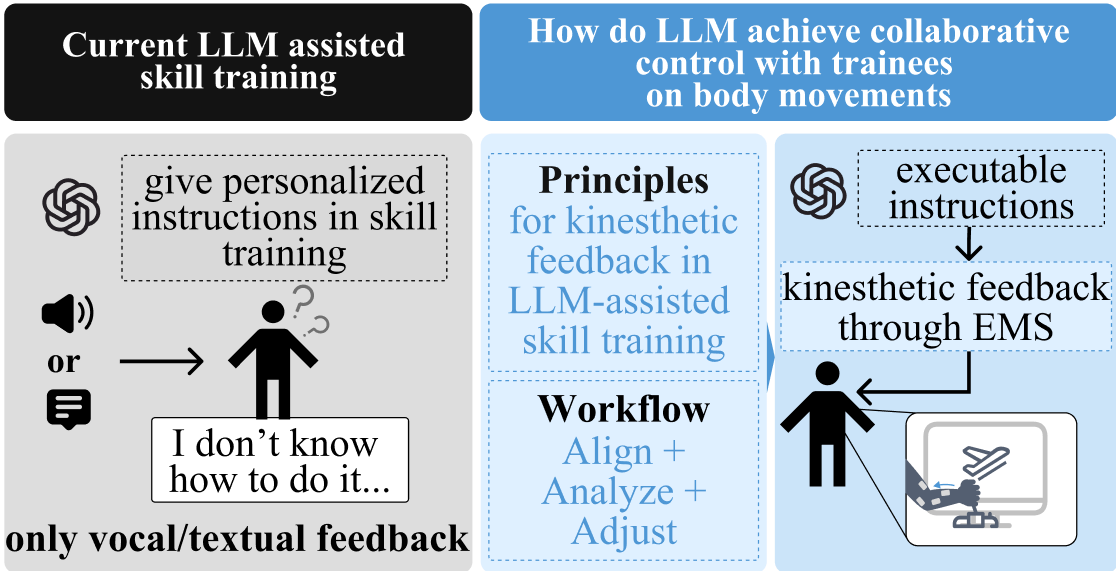}
    \caption{The main contribution of our work to exploring LLM-assisted skill learning with kinesthetic feedback}
\end{figure}

%肢体交互和反馈作为一种有潜力的方式
Despite the strengths, a critical gap remains. In accordance with the language base of LLM, LLM-based systems employ the mainstream way of human-LLM collaboration and assist trainees through vocal and textual feedback \cite{ijcai2024p0869,ijcai2024p193}. This is natural for LLM outputs while leaving the kinesthetic feedback underexplored. These operational skills are inherently physical, requiring hands-on practice and the development of muscle memory through kinesthetic feedback, both for novice and experts \cite{scirobotics}. However, this transformation from vocal to tactile is not straightforward. Vocal instructions, such as ``push forward'', have ambiguities in amplitude and velocity compared to kinesthetic instructions. Also, body control by LLM may lead to resistance even in the context of operational skill learning. How would trainees perform with and react to this kinesthetic feedback? This is the key research question of our article.

%我们的做法
This work focuses on flight as a representative operational skill domain to explore human interaction with LLM-driven kinesthetic assistance. Specifically, we developed FlightAxis \footnote{Supplementary Materials: https://github.com/Z500-RAY/LLM-Driving-EMS-Assistant}, a LLM-supported prototype that utilizes Electrical Muscle Stimulation (EMS) to guide the forearm movement through flight tasks. EMS, by inducing involuntary muscle contractions through targeted electrical stimulation, provides a direct and tangible form of kinesthetic feedback. FlightAxis leverages an LLM trained in flight manuals and expert knowledge to generate personalized guidance, translating textual instructions into EMS patterns that assist trainees in executing flight maneuvers. To the best of our knowledge, it is the first study that combines LLM and EMS for operational skill learning. 

%我们的检验和反馈
A comparative study\footnote{Ethical approval: Organization Ethical Review Committee} was conducted to the effectiveness and user experience of LLM-driven kinesthetic feedback. Participants exhibited positive reactions but also express concerns on specific scenarios, providing insights for design of kinesthetic assistance systems. 
Overall, the contributions include: 
%Contribution
\begin{enumerate}
    \item This study explores the integration of LLM in operational skills learning. LLM and EMS reinforce the learning environment of embodied cognition, making the learning process dynamic and interactive.
    \item This study developed a flight operation skill learning prototype. It employs LLM to align relevant operations, analyze flight status, and generate targeted EMS patterns to adjust arm movements, thereby supporting flight skill learning. 
    \item This study explored human behavior and feelings towards kinesthetic feedback through an empirical study. Our results demonstrate that this approach not only improves operational performance but also enhances trainees' awareness of their own skill gaps. These findings offer valuable insights for designing effective and user-centered kinesthetic training systems.
\end{enumerate}

\section{Related Work}
\subsection{LLM-Assisted Skill Learning}
The application of LLM in skill learning has a solid foundation. Compared to textual manual and recorded video instructions, LLM summarizes operation instructions and enables interactive conversation. This interaction make LLMs ideal for skill learning assistance. For example, Murtaza et al. demonstrated that the use of a ChatGPT-Based Interactive User Manual could improve training outcomes by constructing an flexible learning environment. The trainees could engage in a textual Q\&A with a context-aware manual pre-configured with training scenarios on a simulator screen. Their results showed this approach reduced reaction time for ADAS function activation and increased accuracy \cite{murtaza2024transforming}. Also, Varas et al. reported LLM transcription and summarization of communication among trainees and instructors, thus improving learning outcomes \cite{varas2023innovations}. 

The second advantage of LLM lies in its ability to provide personalized guidance. LLMs monitor trainees' performance and remind criterion whenever needed.
For instance, Kabir et al. proposes an online learning system where LLMs, like ChatGPT, generate personalized practice questions, answers, and feedback based on trainees' knowledge levels, offering a more tailored learning experience \cite{kabir2023llm}. Also, Looi et al. designed a framework of personalized learning affordance enabled by ChatGPT where the tutoring chatbot delivers responses at multiple personalization levels through corresponding prompts \cite{looi2025personalization}. The chatbot effectively identified learners' individual learning gaps and provided personalized support to address them, showcasing LLMs' ability as a training companion.

In general, LLMs have shown the ability to summarize manual of operational skills and give instructions. Inherited by its textual nature, most LLM-supported systems outputs vocal or visual feedback, and have not provide hands-on instructions and kinesthetic feedback for operational skills. Differed from vocal feedback, this direct manipulation of human body meets the nature of physical training, while also requires accurate, short and acceptable outputs.

\subsection{Skill Learning From Kinesthetic Feedback}
Kinesthetic feedback is a widely accepted and essential mode of operational skill learning. Kinesthetic feedback makes full use of human receptors in muscles, skin, tendons, and joints to remind trainees of position, velocity, and force of operations \cite{schostek2009review}.
This feedback helps human locate body parts in space and master the dynamics of movement, even without visual information \cite{pinzon2017skill}. Because operational skills involve body movements and emphasize procedural knowledge, kinesthetic feedback helps trainees refine their actions. For example, providing force and position guidance from robot arm in welding reduced novice welders’ task completion time and improved accuracy \cite{ye2024enhance}. Similarly, multi-directional force feedback and tool-tip tactile sensation significantly reduces errors during robotic surgery, and improves surgeons’ ability to control interaction forces and enhances surgical precision \cite{lim2015role}. Also, the operational skill, which requires development of muscle memory, could largely benefit from continuous repetition. Kinesthetic feedback during operation repetition ensures that learners adjust and refine their technique, helping them develop the physical intuition required to perform the skill autonomously in the future. This has been long validated in surgical training, where repetitive practice of correct techniques helps trainees develop muscle memory, improving skill retention and performance under pressure \cite{buscarini2009training}. However, the way of providing kinesthetic feedback varies across instructors. Trainees might not trust generative systems and need autonomy during learning. For example, feygin et al. reported that overly directive feedback can lead to dependency, reducing a learner’s autonomy and the ability to self-correct \cite{feygin2002haptic}.
Thus, although articles on human instructors have thoroughly discussed feedback strategies, the challenge still, if not more, lies in designing appropriate strategy for LLM driving kinesthetic feedback.

\begin{figure*}
    \centering
    \includegraphics[width=\linewidth]{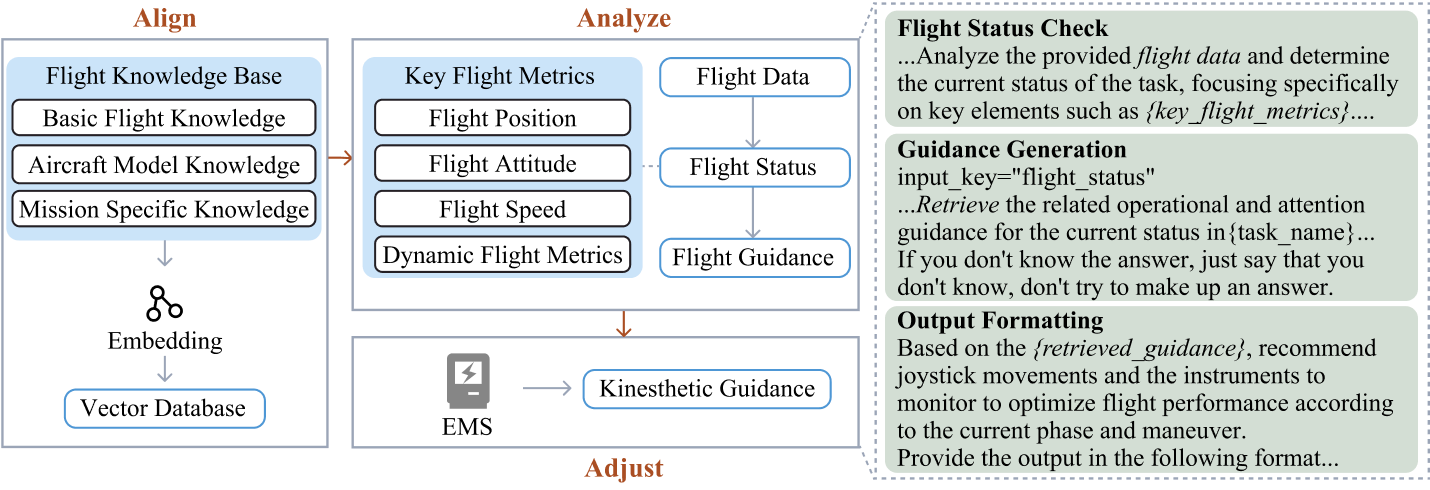}
    \caption{Align-Analyze-Adjust Workflow}
    \label{fig: Workflow}
\end{figure*}
\section{LLM-Supported Kinesthetic Feedback}
In this section, we first adopt principles for kinesthetic feedback in the context of LLM-assisted skill learning.
Considering the benefits and limitations, a system should follow the three principles.
\begin{itemize}
    \item Providing accurate guidance.  A short but accurate guidance helps learners develop an intuitive understanding of the required movements \cite{ye2024enhance}. For instance, in a movement-based task, instead of verbal descriptions like ``move your arm forward'', the system delivers force that guides the learner through the correct motion pathway.
    \item Providing accompanied guidance. Learners need guidance along the training task rather than prior or posterior reflection \cite{buscarini2009training}. 
    \item Maintaining learner control. Kinesthetic feedback serves as a suggestion rather than a constraint, encouraging learners to take actions rather than follow effortlessly. For instance, when operating a control stick, although the system might provide forward-directed kinesthetic feedback, learners maintain the ability to move the stick backward based on their judgment or preference. 
\end{itemize}

To implement kinesthetic feedback in our prototype, we selected Electrical Muscle Stimulation (EMS), a technique that uses an electric current of a specific frequency and intensity to activate muscle fibers, which triggers muscle contractions \cite{PossessedHand,Dexterity}.
EMS can be used to create force feedback for skill learning \cite{FootStriker}, as different stimulation intensity could bring varied effects.
Mild stimulation may only make the user feel lightly touched, strong stimulation induce significant muscle contraction and involuntarily movements \cite{Paired-EMS}.

\section{Align-Analyze-Adjust Workflow}

This section answers ``How we design a LLM supported assistant that provides professional, hand-by-hand guidance''. We developed the ``Align-Analyze-Adjust'' workflow strategy (Figure \ref{fig: Workflow}) to ensure reliable, kinesthetic assistance. Specifically, we ``align'' relevant flight expertise from knowledge base to ensure accurate reference, ``analyze'' flight status to generate precise guidance, and ``adjust'' LLM outputs into kinesthetic feedback of output devices.

\subsection{Task Formulating}
We here chose flight skill learning as a typical task. 
Flight skill learning demands skill on flight procedures, technical skills, and scan patterns. Therefore, there is a significant need for external observation and feedback to optimize learning outcomes \cite{AdoptionVR,AbInitio,operationGap}. 
Flight simulation tools offer self-directed practice but often lack the capability to deliver professional, personalized feedback \cite{McDermott2005,Degani1993345,Rantz2009}.

\subsection{Alignment}
All outputs of LLM should be aligned with standard operation.
Given the vast flight knowledge, we build a flight knowledge base and employ Retrieval Augmented Generation (RAG) technique to provide expertise to LLM.
RAG allows the LLM to tap into the Flight Knowledge Base's wealth of expertise, thereby enhancing the LLM's capacity to generate accurate, contextually rich responses \cite{rag1,rag2}.

We organize the flight expertise into three types:
\begin{itemize}
    \item Basic Flight Knowledge provides the LLM with essential understanding of aerodynamics and flight control.
    \item Aircraft Type Knowledge adapts these fundamentals to the systems and performance profiles of specific aircraft types.
    \item Mission Specific Knowledge incorporates operational details, such as standard operating procedures, flight instrument data, and emergency protocols, empowering the LLM to master complex flight missions with expertise and precision.
\end{itemize}

We used a vector database to store and mark the Flight Knowledge Base files, as a structured and simplified knowledge base can reduce search errors and prevent missing content \cite{li2024virtualcopilotmultimodallarge}, making the search process more efficient and accurate through accessible and hierarchical data organization.

Then we used Facebook AI Similarity Search (FAISS) to manage and search through these vector embeddings \cite{faiss_website}.
Its scalability ensures that even with growing datasets, retrieval remains fast and accurate, which is critical for professional flight operation instruction.

\subsection{Analyses}
\subsubsection{Flight Status Check}
LLM analyze the flight status based on flight training metrics. Given the diversity and volume of flight data, key flight metrics need to be selected to enhance the accuracy of LLMs and to mitigate memory loss caused by excessive context length. 4 key flight training metrics described the status of common flight instrument and flight operations: Flight Position (e.g., altitude relative to sea level), Flight Attitude (e.g., pitch angle, bank angle, magnetic heading angle), Flight Speed (e.g., indicated airspeed, ground speed, vertical velocity), and Dynamic Flight Metrics (e.g., longitudinal acceleration, lateral acceleration, vertical acceleration).
\subsubsection{Guidance Generation}
After checking the flight status, the workflow analyzes the flight status to generate contextually relevant guidance.
% There are two types of guidance: Procedural Guidance to direct the next steps in the flight process, and Corrective Guidance to rectify abnormal flight trends. 
Generated guidance is sorted out and formatted by LLM to translate the guidance for control stick operations into a format acceptable to the EMS device and to extract the names of instruments that require voice feedback. 

\subsection{Adjustment}
Kinesthetic feedback could help initiate an operation and also maintain an operation. Therefore, we first test trainees' reactions to EMS modes and adjust accordingly.
\subsubsection{EMS Guidance Modes} \label{sec:EMS Test}

\begin{figure}
    \centering
    \includegraphics[width=\linewidth]{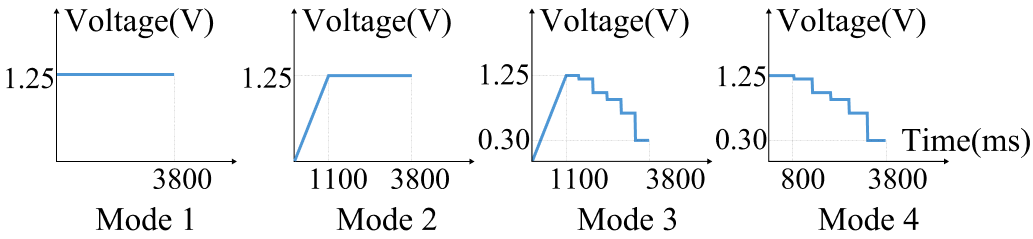}
    \caption{Waveforms of 4 EMS Modes. Time is measured in milliseconds (ms) across all modes.}
    \label{fig: EMS Modes}
\end{figure}

\begin{table*}
    \centering
  \begin{tabular}{cll}
    \toprule
    Item & Question & Character \\
    \midrule
    Q1& Electrical muscle stimulation pushed me in a certain direction. & Sense of EMS direction\\
    Q2& The feedback made my body move on its own. & Autonomy of EMS\\
    Q3 & I was in control of my movements. & Autonomy of human\\
    Q4 & I could associate a movement with the feedback. & Collaboration\\
    Q5 & The feedback did not induce any discomfort. & Comfort level\\
    Q6 & The initiation of feedback did not startle me. & Comfort level\\
    \bottomrule
  \end{tabular}
  \caption{Question to assess the quality of the feedback}
  \label{tab:EMS Table}
\end{table*}

In order to explore the most suitable EMS output mode for pre-start and correction, we conducted a pre-experiment to test 4 EMS modes:
\begin{itemize}
    \item Mode 1 - ``constant strength''
    \item Mode 2 - ``weak-strong'' 
    \item Mode 3 - ``weak-strong-weak'' 
    \item Mode 4 - ``Strong-weak'' 
\end{itemize}

Their waveform charts are shown in Figure \ref{fig: EMS Modes}.

5 participants (2 men and 3 women) took part in the pre-experiment, aged from 19 to 23.
With reference to PossessedHand's procedure \cite{PossessedHand}, we calibrated for each participant prior to the pre-experiment.

During the experiment, participants were tasked with perceiving and responding to four EMS modes. 
They were instructed to hold a control stick tightly with their right arm on a horizontal desktop, using the control stick to execute actions activated by the EMS signals. 
Following each EMS mode performance, participants evaluated the feelings of the EMS mode by completing a questionnaire, with each question scoring on a 100-point scale. 
This questionnaire (Table \ref{tab:EMS Table}), which refers to Faltaous et al. \cite{EMStriker}, was designed to assess the comfort level of different EMS modes. 
After the completion of the tasks, participants were interviewed to gather adjust suggestions.

To assess the sense of control in EMS, we calculated the average score for questions Q1 and Q2. Similarly, to evaluate the comfortable level, we calculated the average score for questions Q5 and Q6.

The sense of control of each group of EMS reached more than 70 points, indicating that the feedback feeling of EMS was overall clear.
Mode 3 had the highest sense of combination of action and feedback (average 81.2), the strongest sense of autonomous control (average 76.8), and the most comfortable sensation (average 74.2). 
Mode 2 had the second highest sense of combination of action and feedback (average 73.8), the second strongest sense of autonomous control (average 74.6), and a stronger sense of EMS direction than Mode 3 (average 79.4). 

For the initiation of operation, it is essential to preserve the user's autonomy while maintaining control. 
In contrast, correcting abnormal operations and maintaining proper procedures requires clear directional feedback and a stronger guiding influence.
Therefore, Mode 3 is selected for the initiation of operation, Mode 2 is selected for the correction of operation.

\subsubsection{EMS Execution}
As the next stage of the flight task approaches, the EMS device initiates the ``pre-start''  (refer to Mode 3 in  \ref{sec:EMS Test}) upon receiving the go-ahead from the preceding part to nudge trainees to prepare themselves. 
This mode functions like a sprinter's starting block or the initial steps in ball sports, providing a guided initiation that enhances control and ease of movement.

During flight operations, maintaining a specific flight status requires constant monitoring and reminders. 
In case of an abnormal flight trend, the EMS, upon receiving information from the preceding part, uses Mode 2 (see \ref{sec:EMS Test}) to guide actions in the direction opposite to the deviation. 
This helps the pilot trainee adjust control inputs and correct the flight path. 

Concurrently, a voice prompt identifies relevant instrument, directing the trainee's attention to the relevant flight parameter for immediate action.

\section{Capability Experiments}

We conducted an experiment to test the feasibility of using RAG-Augmented Guidance Generation part to provide correct professional guidance.

\subsection{Data Collection}
Flight encompasses a wide range of skills that progress from basic to advanced levels. We selected 4 representative tasks-Straight and Level Flight, Normal Takeoff and Climb, Steep Turn, and Deadstick Landing-from the ``Flight Instructor for Airplane Category Airman Certification Standards'' document, whose difficulty increases progressively \cite{FAA2023FlightInstructorAirplane}. These tasks were chosen based on their increasing complexity and their ability to cover essential flight skills across different levels of difficulty.
Each task comprises four test scenarios: two conducted under normal flight conditions and two under abnormal flight conditions.
Normal flight conditions refer to tests conducted under ideal circumstances, such as clear weather and stable flight dynamics. Abnormal conditions include tests with unexpected situations where the flight state deviates from normal standards, such as extreme weather events or unusual flight behaviors.
Microsoft Flight Simulator (MSFS) was used as flight data source, because MSFS has been utilized in the classroom in an effort to develop trainee flight skills \cite{Beckman2011}.
We use Python-Simconnect, a python interface for MSFS2020 using SimConnect to access flight data \cite{odwdinc_python_simconnect}.
Our implemented script retrieves flight metrics from MSFS2020 then standardizes the altitude data and the angular data.
It outputs one piece of data per second in JSON format, ensuring real-time access to the flight training situation.
We processed the flight data for each second through the three-stage workflow, combining the outputs from each stage into a single data record.
This approach allowed us to generate a comprehensive data record for each second of flight task.

\subsection{Experimental Setups}
We input the flight data into the workflow and obtained the outputs of three stages of LLM processing within a complete workflow execution.
The LLM model used is ``GPT-4o-2024-08-06'' from OpenAI \cite{openai2024structuredoutputs}.
We set up three criteria: 
\begin{enumerate}
    \item Completion of all three LLM processing steps.
    \item Generation of output in accordance with the prompt requirements.
    \item Consistency of the output results with the current flight state, including the validity of the reasoning behind the generated action instructions.
\end{enumerate}

Only when the output content of the workflow meets all three conditions, the workflow's processing of these data is recorded as correct.

\subsection{Result}
For each of the four tasks, the following numbers of data records were collected from workflow processing: 390 for Straight and Level Flight, 5773 for Normal Takeoff and Climb, 242 for Steep Turn, and 342 for Deadstick Landing.
This revised statement clearly associates each number with the correct task, providing a precise overview of the data collection for each task.
LLM outputs were evaluated against the criteria.
The judgment accuracy for each task was as follows: Straight and Level Flight at 93.3\%, Normal Takeoff and Climb at 95.5\%, Steep Turn at 91.6\%, and Deadstick Landing at 92.6\%.  The total judgment accuracy among all tasks was 93.2\%.
Specifically, the accuracy under normal flight conditions was 95.6\%, while under abnormal flight conditions it was 92.8\%.

The high accuracy rate indicates that the LLMs performed well in both normal and abnormal flight conditions. This suggests that the models can effectively process flight data and generate reliable outputs.
The slightly lower accuracy under abnormal flight conditions suggests that the models may face greater challenges when dealing with unexpected or complex scenarios.  
This is expected, as abnormal conditions often involve more variability and less predictable conditions.
\section{Empirical Study}
To test the effects of FlightAxis on training performance, the study included 2 training groups: Solo Training Group and Assisted Training Group.
Participants will be randomly assigned to one of the training groups.
Training sessions are conducted within the MSFS platform.
The aircraft selected for training is the Diamond DA40 \cite{diamondaircraft2024}, a model renowned for its widespread use in aviation education.
``Steep Turns'' was chosen as a standard flight training course in the FAA curriculum considering its high frequency in flight operations.

\subsection{Simulator Setup}

\begin{figure}
    \centering
    \includegraphics[width=1\linewidth]{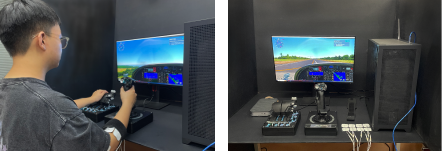}
    \caption{Training Environment}
    \label{fig: training environment}
\end{figure}

The training environment (Figure \ref{fig: training environment}) comprises a computer with MSFS installed, a DELL U2720Q 27-inch LED monitor, a speaker box and a set of Logitech G X56 H.O.T.A.S Throttle and control stick Flight Simulator Game Controller as flight peripherals.

The FlightAxis circuit is integrated into a board. 
It is connected to a computer via a USB cable for data transmission. 
The circuit is linked to the arm through electrode pads, which are securely held in place by an arm loop. 
This design ensures a stable connection between the circuit and the arm.

\subsection{Procedure}
% Figure \ref{fig:Procedure} illustrates the study's procedure.
Before beginning of the experiment, participants completed a pre-questionnaire to provide demographic information as well as their proficiency in flight.
They were randomly assigned to the Assisted Training Group and the Solo Training Group.
After a familiarization Phase including theoretical familiarization and practical familiarization, in the pre-training Phase, participants performed a test task in MSFS, executing one steep turn to the left and one to the right, to establish a baseline performance level.      
During the training Phase, participants in the Assisted Training Group received device calibration and had electrode pads attached to their arms to determine the most suitable current intensity.      
All participants then performed three rounds of steep turn training in MSFS, with each round consisting of one left and one right steep turn.  
In the testing Phase, both groups completed two rounds of testing in MSFS, performing one steep turn to the left and one to the right without any auxiliary devices, while the experimenter timed their task completion. 
Finally,  participants filled out the Flight Skills Operation Learning Experience Questionnaire and underwent an interview to gather their feedback.

\subsection{Participants}
We recruited 36 participants with 20 (56\%) males and 16 (44\%) females, all of whom were healthy and had no conditions that would make EMS unsuitable, such as heart disease, epilepsy, and recent surgical operation.
With a mean age of 21.72 years (\textit{SD} = 2.85), none of the participants have experience of flying or playing flight simulation games. 

\section{Results}
\subsection{Task Completion Quality}
\textit{Altitude maintenance}. For the difference in proportion of altitude falling within the standard interval before and after training, there was a significant difference in the reduction of task completion time between the Solo Training Group (\textit{M} = .21, \textit{SD} = .25) and the Assisted Training Group (\textit{M} = -.02, \textit{SD} = .23) before and after training at the 0.05 significance level (\textit{t}(34)= -2.663, \textit{p}= .012).

\textit{Speed Control} displays no significant difference between the Solo Training Group (M = .12, SD = .39) and the Assisted Training Group (M = .02, SD = .25) at the 0.05 significance level (t(34)= -.964, p= .342).

\textit{Slope Angle Stability} during the hover phase at a 45-degree slope angle  shows no significant difference between the Solo Training Group (\textit{M} = .09, \textit{SD} = .25) and the Assisted Training Group (\textit{M} = .09, \textit{SD} = .22) at the 0.05 significance level (\textit{t}(34)= -.001, \textit{p}= .999).

\textit{Course Angle Accuracy} displays no significant difference in the reduction of course angle deviation between the Solo Training Group (\textit{M} = -16.64, \textit{SD} = 33.45) and the Assisted Training Group (\textit{M} = -8.59, \textit{SD} = 35.28) before and after training at the 0.05 significance level (\textit{t}(34)= -.702, \textit{p}= .488).

\subsection{Task Completion Time}%ok
Independent samples t-test was used to determine the effect of using FlightAxis or not on the reduction in task completion time before and after training.
There was no significant difference in the reduction of task completion time between the Solo Training Group (\textit{M} = -15.62, \textit{SD} = 13.48) and the Assisted Training Group (\textit{M} = -26.40, \textit{SD} = 34.07) before and after training at the 0.05 significance level (\textit{t}(34)= -1.248, \textit{p}= .220).

\subsection{User Experience}

\begin{table}
\centering
  \begin{tabular}{p{3.5cm}p{2cm}p{2cm}}
    \toprule
    \multirow{2}{*}{User Experience} & Solo Training Group & Assisted Training Group \\
                                     &  \textit{Mean}(\textit{SD}) & \textit{Mean}(\textit{SD})\\
    \midrule
    Subjective Task Performance& 4.72(0.90) & 4.33(0.97) \\
    Effort & 4.78(1.35) & 5.44(1.04) \\
    Frustration & 3.00(1.37) & 3.06(1.39) \\
    Immersion & 5.78(0.94) & 5.83(0.71) \\
    \bottomrule
  \end{tabular}
    \caption{User Experience of the Two Groups}
  \label{tab:User Experience of the Two Groups}
\end{table}

For the Assisted Training Group,the prototype's usage comfort level was rated with a mean score of 4.67 (\textit{SD} = 1.00). In terms of process memory assistance, the prototype achieved a mean score of 5.36 (\textit{SD} = 1.36). For its support in operational task learning, the prototype received a mean score of 4.89 (\textit{SD} = 1.33). The prototype's contribution to enhancing situational awareness was rated with a mean score of 5.28 (\textit{SD} = 1.28). In assisting with skill acquisition, the prototype had a mean score of 5.22 (\textit{SD} = 1.27). Its acceptability was rated with a mean score of 5.44 (\textit{SD} = 1.12). Regarding overall satisfaction with the prototype, the mean score was 5.67 (\textit{SD} = 0.94).

Independent samples t-test shows that there was no significant difference in subjective task performance (\textit{t}(34)= -1.250, \textit{p}= .220), effort (\textit{t}(34)= 1.657, \textit{p}= .107), frustration (\textit{t}(34)= .121 \textit{p}= .905), and immersion (\textit{t}(34)= .200, \textit{p}= .843) between the two groups.
Table \ref{tab:User Experience of the Two Groups} shows the detailed statistical results of metrics above. 

Participants in the post interview provided positive feedback on FlightAxis, particularly highlighting its significant assistance for novices during the initial learning phase.  
They have found the prototype system offers clear guidance and identifies issues they may have overlooked.  Feedback from the system lets the user know when to correct their actions and when to proceed to the next step, which leads to significant skill improvement.
The prototype’s feedback combining auditory cues and muscle stimulation, was widely regarded as intuitive and explicit, facilitating a more efficient acquisition of flight skills.      
Participants highlighted that auditory prompts alerted them to issues, while muscle stimulation provided clear guidance for corrective actions, rendering the feedback highly perceptible and actionable.

However, participants also identified several concerns.
They have noted a need for enhanced accuracy in the system's directional control to avoid overcorrections.  
Moreover, although kinesthetic assistance is advantageous for novices, it may overly restrict the actions of more experienced users, thereby potentially hindering their autonomous control and operational fluency.
Overall, the prototype's corrective guidance has been influential, assisting users in refining their actions effectively.

\section{Discussions and Future Work}
We developed a prototype named FlightAxis, which integrates LLM and EMS, to reinforce pilot trainees' flight skills through kinesthetic feedback.

Although it is a short-term training, the trainees achieved significant improvement in altitude maintenance. Overall, the prototype demonstrated positive results in comfort, process memory, operational learning support, and situational awareness. 

In the realm of embodied learning in HCI, physical interaction with the environment is fundamental to learning and understanding. 
Through its ‘pre-start' mode that physically guides trainees' movements, FlightAxis engage the user's kinesthetic sense, aligning with the principles of embodied cognition.
Meanwhile, FlightAxis encourages active user engagement, fostering deeper understanding and adaptability through EMS and voice prompts. 

As a form of human augmentation, FlightAxis highlights the potential of co-adaptive control in designing future AI systems that integrate with the human body, while preserving user autonomy in training. Also, our exploration offers inspirations for AI systems that require close human AI collaboration in complex tasks such as surgery and remote control.
Compared to traditional methods like rule-based systems or conventional machine learning, LLMs with RAG are more flexible and adaptable. Rule-based systems are limited to static tasks due to their fixed rules, while traditional machine learning struggles with new scenarios due to its reliance on labeled data. In contrast, LLMs with RAG combine retrieval and generation to access up-to-date knowledge and provide personalized, context-aware guidance. This leads to more adaptable outputs and facilitates convenient knowledge updates. Future work will compare RAG-augmented LLMs to traditional methods more comprehensively, focusing on guidance quality and user experience, to refine the framework further.

Our experiments show that the LLM's output accuracy is over 90\%, but the randomness and hallucination risks in large language models are still significant concerns. To address these issues, we have standardized the output format using structured prompting techniques and leveraged RAG to reduce hallucinations. Future work could improve model reliability by enhancing data quality through rigorous filtering and validation of external sources, refining RAG strategies, incorporating confidence scoring to filter out unreliable generations, and adopting fine-tuning approaches using high-quality aviation domain datasets to enhance consistency and reduce hallucination rates.

Our prototype ensures safety with the clinically approved EMS device, following protocols for correct electrode pad placement, starting at low intensities, and maintaining moisture on the electrode pads.
Given that the selected tasks are not time-critical, the integration of LLM is appropriate despite seconds of delay. However, highly time-sensitive operations would require local models or other optimization approaches to meet strict real-time requirements.The delay in feedback can cause a sense of lag, which is problematic in scenarios requiring rapid responses. Nonetheless, a moderate delay does not always detract from learning effectiveness. For instance, in scenarios where immediate feedback is less critical, a brief pause for reflection before guidance is provided can foster critical thinking and improve long-term knowledge retention.Additionally, for non-reflexive skill learning, feedback accuracy and relevance may be more important than timing.

The use of non-flight cadet participants potentially led to lower average flight performance, while limited training sessions resulted in less significant learning effects and higher result variability. 
To mitigate this, we recruited college students age-matched with flight cadets and implemented a familiarization Phase combining written materials and verbal instruction. We also conducted two final tests with four steep turns to ensure reliable performance measurement.

Future work could explore various supplementary methods for LLM supported kinesthetic assistance. Other approaches, such as robot arms and haptic feedback can be integrated to enhance the learning experience and provide more adaptive, personalized training solutions.

\section{Conclusion}
This study explored how LLM achieves collaborative control with trainees on body movements by adopting principles for kinesthetic feedback in the context of LLM-assisted skill learning. Based on the principles, we developed the ``Align-Analyze-Adjust'' workflow to customize prompts for real-time feedback on flight skills training, using LLM to process professional procedures and operational specifications while employing EMS to guide the trainee’s arm movements.
The empirical study demonstrated improvement in the quality of flight task, highlighting the device's effectiveness in supporting operational skill development.
User studies show that the LLM-supported kinesthetic assistant led to trainees' significant improvements in altitude maintenance, with positive feedback on comfort, process memory assistance, operational learning support, and enhanced situational awareness.
Our study provides valuable insights and strategies for the integration of LLMs and kinesthetic feedback for skill training, informing future developments in human-AI collaboration systems for complex skill learning.

\section*{Ethical Statement}
This work involved human subjects in its research. All ethical and experimental procedures have been approved by the school's ethical committee.

\section*{Acknowledgements}
This paper is supported by national natural science foundation (62407038) and Aeronautical Science Fund (2023M074076001).

\bibliographystyle{named}
\bibliography{ijcai25}

\end{document}